\documentclass[12pt]{article}
\usepackage{amsmath}
\begin{document}
\begin{center}
\LARGE
\textbf{Spacetime in Everett's
interpretation of quantum mechanics}\\[1cm]
\large
\textbf{Louis Marchildon}\\[0.5cm]
\normalsize
D\'{e}partement de chimie, biochimie
et physique,\\
Universit\'{e} du Qu\'{e}bec,
Trois-Rivi\`{e}res, Qc.\ Canada G9A~5H7\\
email: louis.marchildon$\hspace{0.3em}a\hspace{-0.8em}
\bigcirc$uqtr.ca\\
\end{center}
\medskip
\begin{abstract}
Everett's interpretation of quantum mechanics
was proposed to avoid problems inherent
in the prevailing interpretational frame.
It assumes that quantum mechanics can be
applied to any system and that the state
vector always evolves unitarily. It then
claims that whenever an observable is
measured, all possible results of the
measurement exist. This assertion of
multiplicity has been understood in
many ways by proponents of Everett's
theory. Here we shall illustrate how
different views on multiplicity carry
onto different views on spacetime.
\end{abstract}
\section{Introduction}
Everett's interpretation of quantum
mechanics~\cite{everett1,everett2} was
proposed in a context where challenges to
quantum interpretational orthodoxy were
not well received~\cite{freire}.
At the time, the prevailing view drew both
from the Copenhagen distinction between the
quantum and the classical and from the
Dirac--von Neumann collapse of the state
vector.

Everett's framework attempts to do away
with the collapse of the state vector,
and to correct for a number of
unsatisfactory aspects of the Copenhagen
interpretation.  The latter requires, in
particular, a classical world logically
prior to the quantum world, as well as
observers outside quantum systems under
investigation.  In contradistinction to
this, Everett's framework incorporates the
following characteristics:
\begin{enumerate}
\item The state vector always evolves
unitarily.
\item The `observer' is included in the
quantum description.
\item When a quantum measurement is
performed, all results of the measurement
occur.
\item `Collapse' happens relative to the
observer only.
\item Quantum mechanics applies to the
whole universe.
\end{enumerate}
This general framework was, to some extent,
articulated in Everett's published work
which, however, left a number of questions
only partly answered.  Over the years,
three problems have developed
and received considerable
attention from proponents of (or opponents
to) Everett's approach.  They have to do
with probability, the preferred basis and
multiplicity.
\section{Three problems in Everett's
approach}
\subsection{Probability}
The probability problem in Everett's
approach is twofold.  The first horn
of the problem concerns determinism,
the second one has to do with numerical
values.

As indicated above, according to Everett
the state vector always evolves unitarily,
in accordance with the Schr\"{o}dinger
equation.  So the question is, How can
probability arise if evolution is
completely deterministic?  The answer
given by Everettians is that probability
is not objective.  Rather, it represents an
observer's subjective uncertainty as to
his post-measurement situation.

In quantum mechanics, numerical values of
probability are given by the Born rule.
Everett claimed that the Born rule does
not need to be separately postulated, but
can be derived naturally from the quantum
formalism.  The derivation
turned out to be not so straightforward.
Later investigators attempted to show
that a rational agent who believes he or
she lives in an Everettian universe will
make decisions as if the square amplitude
measure gave chances for outcomes.  This
has given rise to much
debate~\cite{saunders1}, and the issue
is not settled.  
\subsection{The preferred basis}
Consider a two-state quantum system
and an observable $A$ with orthonormal
eigenvectors $|a_1\rangle$ and
$|a_2\rangle$.  A measurement of $A$ will
involve an apparatus in an initial state
$|\alpha_0\rangle$ which, upon interaction
with the quantum system, will evolve to
$|\alpha_1\rangle$ or $|\alpha_2\rangle$,
respectively, if the system is prepared
in $|a_1\rangle$ or $|a_2\rangle$.  

If the quantum system's initial state is
a superposition of $|a_1\rangle$ and
$|a_2\rangle$, the system and apparatus
will evolve as
\begin{equation}
(c_1 |a_1\rangle + c_2 |a_2\rangle)
|\alpha_0\rangle \rightarrow
c_1 |a_1\rangle |\alpha_1\rangle
+ c_2 |a_2\rangle |\alpha_2\rangle .
\label{final}
\end{equation}
But the final state can also be
written as
\begin{equation}
c_1' |a_1'\rangle |\alpha_1'\rangle
+ c_2' |a_2'\rangle |\alpha_2'\rangle ,
\end{equation}
with $|\alpha_1'\rangle$ and
$|\alpha_2'\rangle$ orthogonal
linear combinations of $|\alpha_1\rangle$ and
$|\alpha_2\rangle$ (and $|a_1'\rangle$ and
$|a_2'\rangle$ linear combinations of
$|a_1\rangle$ and $|a_2\rangle$).

Now the question is, Why do we always observe
well-defined macroscopic states $|\alpha_1\rangle$
and $|\alpha_2\rangle$, instead of ill-defined
macroscopic states $|\alpha_1'\rangle$ and
$|\alpha_2'\rangle$?  The most popular
answer is that the property of
decoherence~\cite{schlosshauer}, which is
independent of any interpretational frame,
favors well-defined macroscopic states.
\subsection{Multiplicity}
In Everett's framework, all results of a
measurement occur.  Here the question is,
What does this statement mean from an
ontological point of view?  Everett was not
entirely clear on how to answer this
question, at least
in his published work.  His followers have
developed three distinct types of answer:
\begin{enumerate}
\item Many worlds: The whole universe
splits in different copies.
\item Many minds: Although apparatus
and brain states superpose, consciousness
is well-defined.
\item Reality is identified with patterns
in the universal wave function.
\end{enumerate}
\section{Interpreting quantum mechanics}
Everett's approach provides an interpretation
of quantum mechanics.  Of course, there are
many others~\cite{greenberger}.  But what
does it mean to interpret quantum mechanics?
According to the semantic view of
theories~\cite{fraassen,marchildon1} it
consists in answering the question,
How can the world be for quantum mechanics
to be true?

From this point of view, the problem of
multiplicity is more pressing than the other
two~\cite{marchildon2}.  Indeed simply
postulating the Born rule, as is usually
done, will make probability well defined.
And the preferred basis problem can also
be solved by a specification, adequately
guided by results from decoherence
theory.

The extent of the ontological problem of
multiplicity, gathered from the substantial
literature on Everett's approach, has been
analysed in~\cite{marchildon3}.  The purpose
of this paper is to see how the problem
of multiplicity carries onto
the ontological problem of spacetime.
\section{Many worlds}
As we pointed out above,
Everett's views on multiplicity are
not fully clarified in his published work.
Probably the most straightforward way to
understand multiplicity is to associate
it with a literal split.  That idea was
popularized by DeWitt~\cite{dewitt}:
\begin{quote}
This universe is constantly splitting into
a stupendous number of branches, all resulting
from the measurementlike interactions between
its myriads of components. Moreover, every
quantum transition taking place on every star,
in every galaxy, in every remote corner of
the universe is splitting our local world
on earth into myriads of copies of itself.
\end{quote}
This quote suggests that a split occurs
everytime an interaction produces
entanglement, whether or not macroscopic
objects are involved.  Everett, however,
introduced multiplicity only in contexts
where something like a macroscopic apparatus
performs a measurement.  Considerations
will henceforth be restricted to such
situations.

Healey~\cite{healey} was the first to
formalize the consequences of splitting on
the nature of space.  He first considered
the possibility that systems split into~$n$
copies, in usual ordinary space.  But then,
so he argues, mass-energy would be multiplied
by~$n$ and we would presumably be aware of
the overcrowding of space.  Accordingly,
the split can become acceptable in two
different ways:
\begin{enumerate}
\item The physical systems do not split,
only their states do.
\item Not only systems, but space itself
splits.  The resulting systems may be viewed
as living in a higher-dimensional manifold.
\end{enumerate}
The first way anticipates multiplicity
viewed as decoherent sectors of the wave
function, which we will examine later.
The second way is perhaps the easiest
one to visualize.  It involves multiple
copies of spacetime, all of them multiplying
further upon each measurement interaction.
It does, however, raise
a number of questions usually associated
with state vector collapse: When, in the
measurement process, does the split occur?
Does the split occur on an equal-time
hypersurface, or on a light cone (so as
to be more consonant with special relativity)?
What are the precise conditions that define
a measurement?

Related to splitting is the intriguing
question of recombination.  If evolution
is unitary, measurements can in principle
be undone.  Just like worlds, multiple
copies of spacetime should then recombine.
This, however, can be avoided by means
of bifurcation, an alternative to splitting
introduced by Deutsch~\cite{deutsch1}.
The idea is that there are infinitely
many worlds (or spaces) at any time.
Their number neither increases nor
decreases. In measurement contexts the set
of all spaces is partitioned in as many
subsets as there are possible measurement
results.  In this case the multiplicity of
the whole universe does not change, nor
does the complete spacetime arena.
\section{Many minds}
Albert and Loewer~\cite{albert} gave the
first full-fledged formulation of the
idea that the split involves the mind
rather than the world.
In the many-minds view, every observer has
associated with it an infinite set of minds.
Minds are associated with brain states
but are not subject to superposition.

Many-minds approaches usually involve a
single spacetime, in which different
experiences coexist.  But according to
Lockwood~\cite{lockwood},
\begin{quote}
the fact of a physical system's being in a
superposition, with respect to some set of
[consciousness] basis vectors, is to be
understood as the system's having a
\emph{dimension} in addition to those of
time and space.
\end{quote}

Just as one's mind is usually thought
as being wholly present at different
times when it has different experiences,
it can also be thought, according to
Lockwood, as wholly present in each
different experiences it can have at a
single time.  These experiences can be
viewed as lying on an axis orthogonal
to the time axis.  Spacetime is therefore
enlarged, but by a dimensionality that
is neither spatial nor temporal.
Lockwood does not speculate on how the
spacetime metric can connect to the
additional dimension.
\section{Patterns in the wave function}
Decoherence theory is an important building
block of an approach to Everett different
from many worlds and many minds.  It is
mainly connected with the names of Gell-Mann
and Hartle~\cite{gellmann},
Saunders~\cite{saunders2} and
Wallace~\cite{wallace1,wallace2} (although
Gell-Mann and Hartle favor one world in the
end).

Wallace~\cite{wallace1} identifies real
structures with stable patterns in the
universal quantum state:
\begin{quote}
My claim is instead that the emergence
of a classical world from quantum
mechanics is to be understood in terms
of the emergence from the theory of certain
sorts of structures and patterns, and that this
means that we have no need (as well as no
hope!) of the precision which Kent and others
here demand.
\end{quote}
At the beginning of an experiment,
there is one apparatus pattern in the
universal quantum state.  This, according
to Wallace, means that there is one
apparatus.  At the end of the experiment,
the universal quantum state, represented
on the right-hand side of~(\ref{final}),
contains two distinct apparatus patterns.
Therefore there are now two apparatus.
It doesn't make sense, according to
Wallace, to ask questions about the
apparatus in the very short decoherence
timescale leading from one pattern to
two patterns.  Nor does the existence
of two distinct patterns cause problems
for, as one can see in Schr\"{o}dinger's
cat imagery:
\begin{quote}
If A and B are to be `live cat' and `dead cat'
then [the relevant microscopic properties]
P and Q will be described by statements
about the state vector which (expressed in a
position basis) will concern the wave-function's
amplitude in vastly separated regions $R_P$ and
$R_Q$ of configuration space, and there will
be no contradiction between these statements.
\end{quote}

So the live cat and the dead cat occupy
different regions in configuration space.
But how do they
project in three-dimensional space?
In classical theory, two different cats
not only occupy different regions in
configuration space, but also different
regions in three-space.  That is, their
projections from configuration space to
three-space do not overlap.

This, however, is not so with Wallace's
patterns. Projected in three-space, the
two cats may literally overlap.  How can
one understand this?

One possible answer is to assume that
the live cat and the dead cat don't project
from configuration space to the same
three-space~\cite{baccia}.  This means,
for instance, that there is an
added parameter, or another dimension,
introduced to distinguish different
three-spaces from each other.

This, however, is not the answer that
Wallace prefers.  According to him, there
is only one three-space into which both
patterns project.  This implies that the
live cat and the dead cat are, so to
speak, ghostlike to each other.  In the
words of Allori \emph{et al.}~\cite{allori},
``[t]he two cats are \mbox{[\ldots]}
reciprocally transparent.''

Do we have to choose between one
three-space and many three-spaces?
Investigating macroscopic ontology in
Everettian quantum mechanics,
Wilson~\cite{wilson} suggests
that we may not:
\begin{quote}
[T]he `spacetime' of the quantum mechanics
and quantum field theory formalism,
in terms of which branches are defined,
is not the same as the `spacetimes' of
macroscopic worlds. The former `spacetime'
is a single entity common to multiple
branches, while each of the latter
`spacetimes' is tied to a particular
macroscopic course of events.
\end{quote}
But Wilson also draws an analogy between
bifurcation and consistent histories,
each history being located in a distinct
spacetime.
\section{Discussion}
We have shown elsewhere~\cite{marchildon3}
that there is a wide spectrum of opinions,
among adherents to Everett's approach,
on the nature of Everettian multiplicity.
This, as we have just seen,
translates into different views of
space or spacetime, which fall into
two broad categories:
\begin{enumerate}
\item Space (or spacetime) genuinely splits
upon measurement of a quantum observable.
\item Space (and spacetime) don't split.
\end{enumerate}
To some extent, this alternative occurs in
all three main approaches to multiplicity,
namely, many worlds, many minds and
decoherent sectors of the wave function.

If space genuinely splits, the problems
traditionally connected with state vector
collapse seem to be carried along the way.
This does not mean that Everett's approach
has to be rejected, but it invites
Everettians to make their views
substantially sharper.

If space doesn't split, we are confronted
with various copies of macroscopic systems
occupying the same spatial arena.  The
problem is, How can these systems
not interact?  In other words, How is the
macroscopic multiplicity reconciled with
the quantum field theory of interacting
constituents?  Wilson acknowledges the
problem when he introduces two different
notions of spacetime.  The solution,
however, remains to be implemented.
\section{Conclusion}
Interpreting quantum mechanics consists
in answering the question, How can the
world be for quantum mechanics to be true?
Everett's approach is one possible answer
to the question.  Every answer currently
considered has problems of its own.  We
conclude by pointing out three problems
that are perhaps more specific to Everett's
theory.

The first one is that it is not sharply
defined.  This has been documented here
with respect to spacetime, and
in~\cite{marchildon3} with respect to
multiplicity.  This contrasts with the
de Broglie--Bohm approach which, at least
in the nonrelativistic case, is rather
well defined.  Other interpretations do
carry some measure of indefiniteness,
for instance the Copenhagen distinction
between the quantum and the classical.
But the diversity of views among Everett's
adherents is particularly striking.

The second one can be considered either a
problem or a strength, depending on one's
point of view.  It consists in the fact
that Everett's approach is highly dependent
on the exact validity of quantum mechanics.
Everett's many worlds, so it seems, won't
survive the smallest nonlinear term to the
Schr\"{o}dinger equation.  By contrast,
again, the de Broglie--Bohm approach is
highly adaptable to changes in the formalism
of quantum mechanics~\cite{valentini}.

The third problem has to do with
Everett's extraordinary ontology.
True, science has taught us that common
sense is not always the best of guides.
Yet, as in experimental investigations,
``Extraordinary claims require
extraordinary evidence.''  Many critics
of Everett's approach believe that such
evidence has not convincingly been
put forward by Everett's supporters.
\section*{Acknowledgements}
I am grateful to the Natural Sciences and
Engineering Research Council of Canada
for financial support.
\end{document}